# Lennard Jones Token: a blockchain solution to scientific data curation

Brian H. Lee and Alejandro Strachan[a)]

Author affiliations
[1]*School of Materials Engineering and Birck Nanotechnology Center, Purdue University, West Lafayette, Indiana 47907, USA*

Author email
[a)]Author to whom correspondence should be addressed: strachan@purdue.edu

## Abstract

Data science and artificial intelligence have become an indispensable part of scientific research. While such methods rely on high-quality and large quantities of machine-readable scientific data, the current scientific data infrastructure faces significant challenges that limit effective data curation and sharing. These challenges include insufficient return on investment for researchers to share quality data, logistical difficulties in maintaining long-term data repositories, and the absence of standardized methods for evaluating the relative importance of various datasets. To address these issues, this paper presents the Lennard Jones Token, a blockchain-based proof-of-concept solution implemented on the Ethereum network. The token system incentivizes users to submit optimized structures of Lennard Jones particles by offering token rewards, while also charging for access to these valuable structures. Utilizing smart contracts, the system automates the evaluation of submitted data, ensuring that only structures with energies lower than those in the existing database for a given cluster size are rewarded. The paper explores the details of the Lennard Jones Token as a proof of concept and proposes future blockchain-based tokens aimed at enhancing the curation and sharing of scientific data.

# 1. Introduction

Advances in computational power and algorithms have elevated the importance of machine learning (ML) in the physical sciences and engineering. Findable, accessible, interoperable, and reusable (FAIR) data[1] are critical to realizing the enormous potential of these tools. In the field of materials science and chemistry, several infrastructure efforts have been launched to address this gap.[2–9] Despite these efforts, the vast majority of scientific data collected globally remains unavailable, hindering innovation. For example, while the most extensive database for battery materials contains more than $10^4$ crystal structures, cycling performance, and conductivity is known for an order of magnitude fewer cases.[10] For a more niche topic such as energetic materials, the largest datasets consist of $O(100)$ data points, and publications often lack key information (e.g. microstructural information),[11–13] undermining the effectiveness of modern machine learning models. Other disciplines face similar challenges of FAIR data scarcity.[14,15] While various ML methods are being developed to compensate for the scarcity of data,[16] the scientific community stands to benefit from the expansion of scientific data.

The universal adoption of FAIR principles and scientific data expansion is hampered by several technical and cultural obstacles. First, while sharing scientific data following FAIR[1] principles remains costly in terms of time and labor, incentives are limited. Data-share mandates,[17] community efforts,[18] and journal requirements incentivize the supply and not the demand, encouraging meeting external criteria instead of competitively sharing data demanded by data users. This could lead to suboptimal improvement in scientific databases. Second, the sustainability of scientific data gateways remains a challenge.[19,20] Widely used platforms like the Materials Project,[3] NOMAD,[21] and nanoHUB[22] are supported mostly by government agencies and their long-term sustainability depends on continuing centralized funding, which is not necessarily guaranteed. While private organizations may provide such services, the continued accessibility of data is not guaranteed as evidenced by the decommissioning of platforms like Open Citrination.[23] Lastly, methods to evaluate the relative value of data are often absent, complicating the process of appropriately rewarding data submissions and prioritizing the dissemination of useful data.

Blockchain technology offers key solutions to these challenges in scientific data curation. First, blockchain-based tokens can serve as incentives for researchers to upload and share quality data, where the value of the tokens are determined by users that pay for access to the data via tokens. Second, the decentralized nature of blockchain allows for secure data storage solutions. In addition, smart contracts can automate the process of data evaluation. As a practical demonstration of these capabilities, we introduce the Lennard Jones Token (LJT). Implemented on the Ethereum network, this token rewards users for submitting optimized structures of Lennard Jones particles while charging them to gain access to the existing optimized structures in the database. For such a system, the value of a token is directly related to the value of the data and the querying system provided by the smart contract. Only structures with energies lower than those already in the database are rewarded, ensuring an ever-improving repository of valuable data. Such criteria for data submission act as a consensus algorithm that rewards the computational work of miners without a need for central authority. The proposed framework is designed to harness greed for advancement of science. For example, steering a fraction of computational resources spent on proof-of-work for Bitcoin[24] toward projects such as Folding@home[25] can significantly facilitate the expansion of scientific databases. This paper delves into the details of the Lennard Jones Token and proposes additional tokens that could further motivate the curation of scientific data.

## 2. Background

*2.1 Blockchain*

Blockchain, first proposed by David Chaum[26] and popularized by Satoshi Nakamoto with Bitcoin,[27] is a decentralized and distributed ledger designed to record data across multiple compute nodes. Data in a blockchain are recorded in blocks that are linked with other blocks through cryptographic hashes and shared by all participating nodes. Such data structure ensures the transparency and immutability of data as malicious tempering of a block requires modification of all subsequent blocks. Taking advantage of these characteristics, various applications such as cryptocurrency,[27,28] supply chain management,[29] healthcare,[30] and research[31] have adopted blockchain technology. Additional information about blockchains and online code is available in Ref [32].

*2.2 Cryptocurrency tokens*

Cryptocurrency tokens are virtual assets within a particular blockchain ecosystem. Unlike primary cryptocurrencies such as Bitcoin or Ethereum, which function mainly as a decentralized currency for their respective blockchains, tokens are built on top of existing blockchain platforms using standardized protocols. They can serve as a form of currency that provides utility or governance in a given platform. In Ethereum token standards, tokens are classified into fungible (FT) and non-fungible tokens (NFT). Fungible tokens, generally designed with ERC-20 or ERC-777 standards, are divisible and interchangeable with other tokens. NFTs following ERC-777 standards are not-interchangeable and unique.

*2.3 Molecular or crystal structure optimization*

A key research topic in both molecular biology and material science is predicting optimal structures of molecules, nanomaterials, and bulk systems as their structures play pivotal roles in determining their properties and functionality. Because the molecules tend to reside in energetically stable configurations following the Boltzmann distribution, molecular structure determination involves obtaining the global and local minimum energy configurations of a molecule. Various methods, including simulated annealing,[33,34] basin-hopping optimization,[35] advanced Monte Carlo simulations[36,37] including evolutionary methods,[38] and exhaustive crystal structure exploration[3] have been employed to optimize the geometry of molecules. For LJT, optimized structures of clusters of Lennard Jones particles are demanded. While global energy minimum configurations for LJ particles are already reported in literature,[35] methods of obtaining such structures are considered important as they can be applied to more complex and practical cases such as charged,[39] binary,[40] or anisotropic[37,41] particles as well as clusters at an interface.[42,43]

## 3. Lennard Jones Token

### 3.1 Smart contract architecture

Lennard Jones Token (LJT) is a fungible token following the ERC-20 standard on Sepolia Testnet (contract address: 0x717a61B4F194ACA49DE4fBAA71E083C5660668E5). The purpose of the token is to reward miners for submitting energetically optimized Lennard Jones (LJ) particle clusters while making this data accessible to users in exchange for LJTs. In addition to the standard methods available for ERC-20, the smart contract consists of the following additional functions that will be explained in this subsection: *mineToken*(), *calcEnergy*(), *viewData*(), *gainAccess*(), *setExchangeRate*(), *buyToken*(), *viewTopRate*(), and *viewTopBalance*().

The smart contract stores two variables, *energyDB* and *positionDB*, on the chain. *energyDB* maps a uint value for the number of particles ($N$) in the cluster to an int value for the energy of the cluster. *positionDB* maps $N$ to an uint array of particle configurations. For this prototypical smart contract, the clusters are limited to the size of 2 to 50 particles. The initial values of the energies and positions are set to simple cubic lattices of LJ particles by the owner of the contract.

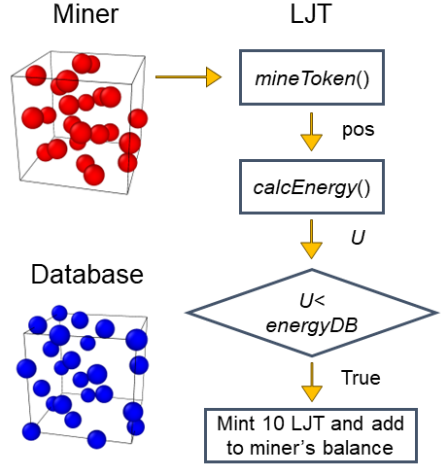

Figure 1. Schematic for mining Lennard Jones Token. A miner sends positions of LJ clusters to the smart contract with *mineToken*() function. The smart contract calculates the potential energy of the cluster with *calcEnergy*() function. If the energy is lower than the energy saved on the database, the miner's address is rewarded 10 LJT and the database is updated with the new cluster.

The schematic for mining energetically optimized LJ particle cluster is given in Fig. 1. A miner can submit configuration of LJ particle clusters as an uint array via *mineToken*() function. The *mineToken*() function internally names the input array as *pos* and calls *calcEnergy*() function with *pos* as input. *calcEnergy*() function calculates the energy of the particle cluster by the standard LJ potential formula:

$$U = \sum_{i=0}^{N-2} \sum_{j=i+1}^{N-1} 4\varepsilon \left[ \left(\frac{\sigma}{r_{ij}}\right)^{12} - \left(\frac{\sigma}{r_{ij}}\right)^{6} \right]$$

, where $U, N, \varepsilon$ and $\sigma$ are the potential energy, number of particles in the *pos* array, the energy and distance scales, respectively. $r_{ij}$ is the distance between particles $i$ and $j$. The $\varepsilon$ and $\sigma$ are set to 1. The *calcEnergy*() function returns $U$. If this value is less than a hard-coded percentage ($\Delta$) of the stored energy for a particle of size $N$, the *energyDB* and *positionDB* variables are updated by the miner's input values, and the miner is rewarded $\rho$ tokens. The parameter $\Delta$ should be set to a value that would prevent malicious submission of nearly equivalent configurations. In this smart contract, $\Delta$ and $\rho$ are set to 3% and 10 tokens, respectively.

A user can gain access to the stored data with *gainAccess*() function. The function accepts a uint variable $N$ as input. If the user's balance of token is greater than 1, the user's balance is reduced by 1 token and the user gains access to mapping of $N$ in *energyDB* and *positionDB*. The balance of miner who contributed the data is increased by 1 token from this exchange. Once access is granted, the user can view the data by calling *viewData*() function.

A user can buy tokens from miners or the owner of the contract by calling *buyToken*() function with desired amount of Ethereum. The amount of LJT corresponding to the exchange rate of the seller is transferred. The miners can set their exchange rate by *setExchangeRate*() function. We limit the profit of the contract owner through greed-limiting-algorithm that sets a minimum exchange rate of the contract owner. If the miner does not define their exchange rate, it is set equal to the minimum exchange rate of the contract owner. The buyer can view the top 10 addresses with largest token balance and minimum exchange rates by calling *viewTopRate*() and *viewTopBalance*() functions. The code for the LJT is shared in github.[44]

## 3.2 Decentralized app for Lennard Jones Token

A decentralized app (dapp) for Lennard Jones Token was developed to interact with the LJT smart contract as shown in Fig. 2. This dapp allows users to connect their wallets with the 'CONNECT WALLET' button and view their ETH, LJT balance, the data access they have, and their exchange rates with 'UPDATE BALANCE' button (Fig. 2a). The addresses for top balances and rates are queried with 'TOP BALANCE/RATE' button while 'SET YOUR RATE' and 'TRADE' buttons allow the user to set their exchange rates and trade LJT with selected addresses (Fig. 2b).

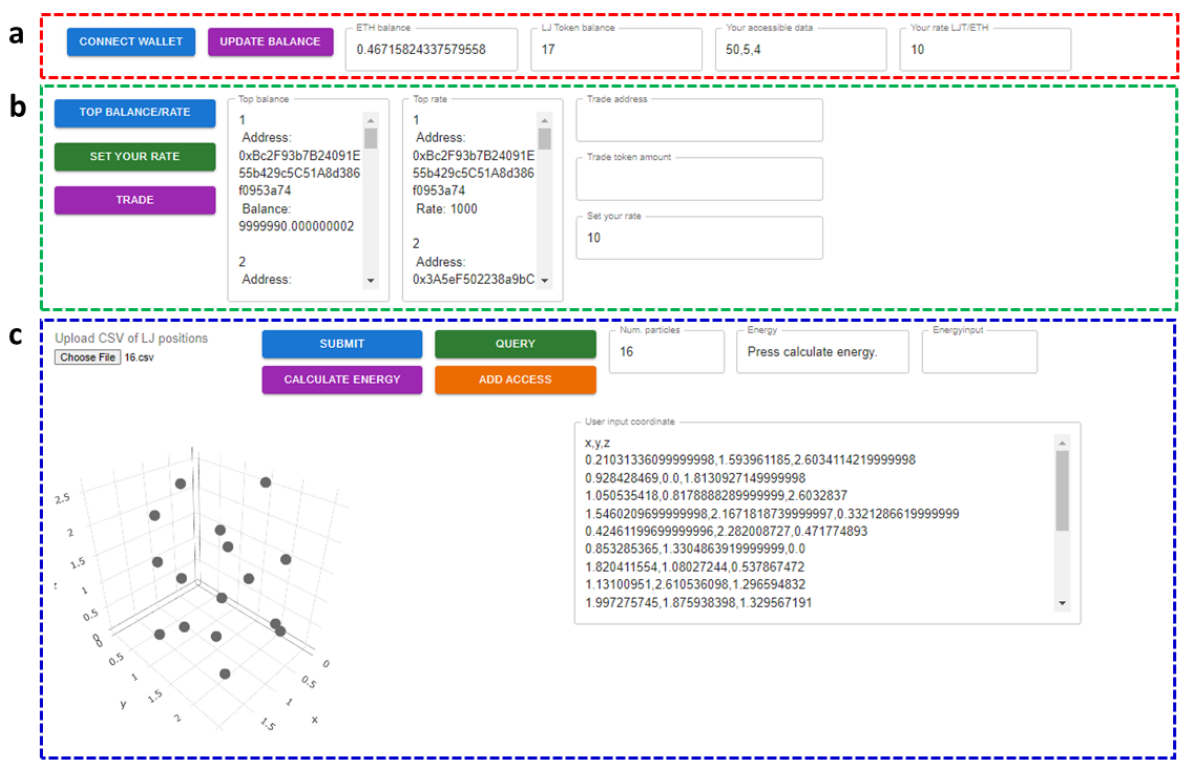

Figure 2. Frontend dapp for interacting with Lennard Jones Token smart contract. (a) Users can connect their Metamask wallets ("CONNECT WALLET") and view their balance of ETH and LJT ("UPDATE BALANCE"). (b) Users can query the balance and exchange rates of others ("TOP BALANCE/RATE"). Users can set their own exchange rates between ETH and LJT ("SET YOUR RATE"). Users can trade their ETH and LJT with others based on the exchange rates set by the others ("TRADE"). (c) LJ cluster position data files can be uploaded. With "SUBMIT," user can call *mineToken*() function to submit the uploaded cluster data to mine LJT. "CALCULATE ENERGY" calls *calcEnergy*() function to calculate the energy of the cluster. "Query" calls *viewData*() to see the position and energy of the cluster with number of particles indicated on Num. particles text field. "ADD ACCESS" gives user access of the data with cluster size indicated on Num. particles text field.

The mining and data access is handled by buttons in Fig. 2c. When a user uploads a csv file with positions of the LJ particles, the particles are plotted and the data is displayed on the panels below the buttons. The 'CALCULATE ENERGY' button calls *calcEnergy*() function with the data from the user's csv file as input. The energy is displayed on the text field 'Energy.' The 'SUBMIT' button calls *mineToken*() function with the csv data. 'QUERY' button calls *viewData*() function with the value in the 'Num. particle' text field as input. 'ADD ACCESS' button calls *gainAccess*() function with the value in the 'Num. particle' text field as input.

## 4. Extension to additional scientific tokens

The LJT was created as a proof-of-concept example of blockchain tokens that can facilitate scientific data curation without expectation for practicality, as optimal structures of LJ particle clusters are already known.[35] However, the framework for this token can be extended to any scientific data for which the scientific value of the data can be objectively evaluated without a central authority. A possible extension of LJT would be to apply the same smart contract architecture for biological molecule simulation trajectories. Understanding the structure and dynamics of proteins is crucial for biomedical research. For example, during the Covid pandemic, a large number of people volunteered their compute resources to simulate the SARS-CoV-2 spike proteins, where the resulting molecular trajectories revealed numerous viable targets for antivirals.[45] While the SARS-CoV-2 has attracted significant societal interest and achieved exascale distributed computing from volunteers, most projects do not command such interest. If a framework like our token can induce even a fraction of the compute resources used for mining bitcoin to be used to simulate protein trajectories, the benefits to the scientific community and society in general would be significant.

For biological molecules, the evaluation criteria for the value of data needs to be changed compared to LJT as existence of transition and metastable states indicates that minimizations of energies are not the only relevant criteria. We propose two possible approaches. First, a decentralized approach can be taken where the energy and the visualization of protein structure is freely available as NFT but users are required to pay tokens to access the numerical data of the molecule trajectory. Second approach is a centralized approach in which trusted experts evaluate the value/price of molecular trajectories depending on the structural diversity and energy as well as importance of the molecule. Other approaches, such as weighting the importance of data by the number of people affected by the molecule, i.e. viral infections, can be considered as well.

Any scientific data whose value can be evaluated by smart contract can benefit from our approach. Additional examples include: 1) binding affinity of drugs, 2) crystal structures of materials whose stability is evaluated using a convex hull, 3) relaxed structures of polymers.

## 5. Conclusions

In this paper, we proposed and implemented a blockchain infrastructure as a solution to the challenges of scientific data sharing, curation, and long terms storage. The blockchain technology, with its inherent properties such as data transparency, immutability, as well as its ability to create value in a decentralized manner, offers a unique paradigm for scientific data compared to traditional cyber-infrastructure that relies on benevolence of the participants for data submission and government agencies or corporations for sustainability. Our exploration of the Lennard Jones Token (LJT) on the Ethereum network showcased how blockchain-based tokens can incentivize quality data submission and storage. The unique design of LJT, which rewards only the submission of superior quality data, demonstrates a decentralized consensus algorithm that continuously refines the databases. We believe that future implementations of our approach to types of data that are considered more valuable to the scientific community, such as trajectory of biomolecules or drug binding affinities, will be important for widespread adoption of our approach.

## Code availability

The codes associated with this work are available in: https://github.com/leebhbrian/LennardJonesToken.[44]

## Acknowledgments

The authors acknowledge the support of Purdue University. AS also acknowledges support from the US National Science Foundation, FAIROS program, OAC 2226418.

## Author declarations

The authors have no conflicts to disclose.